\documentclass{article}
\usepackage{multirow}
\usepackage{subcaption}
\usepackage{spconf,amsmath,graphicx}
\usepackage{amsfonts}

\usepackage{xcolor}
\usepackage{wrapfig}
\usepackage{cite}
\usepackage{booktabs}
\usepackage{threeparttable}
\usepackage{colortbl}
\usepackage{setspace}
\usepackage{cite}
\usepackage{booktabs}
\usepackage{threeparttable}
\usepackage{hyperref}
\usepackage{makecell}
\usepackage{bm}
\usepackage{amsmath}
\hypersetup{hypertex=true,
            colorlinks=true,
            linkcolor=black,
            anchorcolor=black,
            citecolor=black}
\usepackage{pdfpages}
\usepackage{tabularx}
\usepackage{soul}
\usepackage[font=small,labelsep=none]{caption}
\usepackage{spconf,amsmath,graphicx}

\title{Adaptive  speech emotion representation learning based on dynamic graph}
\name{Yingxue Gao \qquad Huan Zhao$^*$ \qquad Zixing Zhang$^*$ 
\thanks{$^\star$ Corresponding authors: H. Zhao and Z. Zhang (email: \{hzhao, zixingzhang\}@hnu.edu.cn). The work was funded by the National Science Foundation of China under Grant Number 62076092.}
}
\address{College of Computer Science and Electronic Engineering, Hunan University, China}
\begin{document}
\ninept
\maketitle
\begin{abstract}
Graph representation learning has become a hot research topic due to its powerful nonlinear fitting capability in extracting representative node embeddings. However, for sequential data such as speech signals, most traditional methods merely focus on the static graph created within a sequence, and largely overlook the intrinsic evolving patterns of these data. This may reduce the efficiency of graph representation learning for sequential data. For this reason, we propose an adaptive graph representation learning method based on dynamically evolved graphs, which are consecutively constructed on a series of subsequences segmented by a sliding window. In doing this, it is better to capture local and global context information within a long sequence. Moreover, we introduce a weighted approach to update the node representation rather than the conventional average one, where the weights are calculated by a novel matrix computation based on the degree of neighboring nodes. Finally, we construct a learnable graph convolutional layer that combines the graph structure loss and classification loss to optimize the graph structure. To verify the effectiveness of the proposed method, we conducted experiments for speech emotion recognition on the IEMOCAP and RAVDESS datasets. Experimental results show that the proposed method outperforms the latest (non-)graph-based models.

\end{abstract}
%

\section{Introduction}
\label{sec:intro}

Graph representation learning has demonstrated tremendous promise due to its powerful capability of mining graph structure information and data relationships \cite{DBLP:conf/icassp/HuHWJM22}. 
Graph convolutional network (GCN), a concrete and popular implementation in graph representation learning, has been widely used since it fully considers the relationships between target nodes and neighboring nodes to learn efficient node representations. For example, Compact-GCN \cite{DBLP:conf/icassp/ShirianG21} constructs a lightweight GCN architecture, which can perform accurate graph convolution for speech emotion recognition (SER).
To model dynamic data, L-GrIN \cite{DBLP:journals/tmm/ShirianTG22} proposes a learnable graph structure, which is designed to adapt across modalities.

Despite great progress made in graph representation learning for SER, they primarily focused on a static graph constructed on an entire utterance. This may fail to capture the trivial variation of emotion in a small region. Moreover, for most previous methods, the dominant paradigm to update the node representations is by averaging the information of neighboring nodes, without considering the importance of different neighboring nodes. Although some efforts have been made to explore a weighted average approach by considering the attention mechanism~\cite{DBLP:journals/tcbb/ZhaoLYNZ23,DBLP:conf/icassp/WuLZGT22}, the calculation process is time-consuming and highly computationally complex.

To address these shortcomings, this paper proposes an adaptive graph representation learning model based on dynamically evolved graphs. Specifically, we consecutively construct the graphs on a set of subsequences segmented by a sliding window, where each node of the graph corresponds to a frame (frame-to-node), and extract the feature vector of the frame as its node representation.
Then, the node representation is updated by our proposed matrix calculation method. Finally, we construct a learnable graph convolutional layer to optimize the graph structure. The contributions of this paper are summarized as follows:
\begin{itemize}
\item We introduce a weighted method to update the node representations rather than the traditional methods of averaging the information of neighboring nodes, where the weights are calculated by the proposed matrix computation based on the degree of neighboring nodes. 
\item We construct a learnable graph convolutional layer that combines the graph structure loss and classification loss to jointly optimize the graph structure during the training. 
\item Experimental results show that our SER model has better performance than the state-of-the-art (SOTA) graph-based networks and several widely used non-graph-based models on the IEMOCAP and RAVDESS datasets.
\end{itemize}

\section{Related work}

There are two kinds of graph representation learning that are used to deal with static graphs and dynamic graphs.
For the former, these methods focus on mining the connectivity of graphs. A well-known connectivity way is to connect the first-order or second-order neighboring nodes \cite{DBLP:journals/tmm/ShirianTG22}, which provide the structure information of a graph at different levels. Commonly used approaches, including random walk (DeepWalk \cite{DBLP:conf/kdd/PerozziAS14} and Node2Vec \cite{DBLP:conf/kdd/GroverL16}), graph convolution (GCNs \cite{DBLP:conf/iclr/KipfW17}), sampling (GraphSAGE \cite{DBLP:conf/nips/HamiltonYL17} and GraphSAINT \cite{DBLP:conf/iclr/ZengZSKP20}), non-negative matrix decomposition (M-NMF  \cite{DBLP:conf/aaai/WangCWP0Y17}), and attention mechanism (GAT \cite{DBLP:conf/iclr/VelickovicCCRLB18}). However, these methods largely ignore the evolution of graph structures and the temporal properties of graphs.

Recently, several dynamic representation learning approaches have been proposed. Specifically, to dig deeper into the local structure of the graph, StudentLSP \cite{DBLP:conf/cvpr/YangQSTW20} utilizes a knowledge distillation method to learn the node representations. 
DySAT \cite{DBLP:conf/wsdm/SankarWGZY20} learns node representations with the help of a self-attention mechanism by modeling both neighboring nodes and temporal attributes.
FADGC \cite{DBLP:conf/icassp/WuLZGT22} captures the temporal properties of dynamic graphs by using a fine-grained attention mechanism to focus on the node changes.  
In addition, to learn the dynamic graph representations of a set of nodes, EvolveGCN \cite{DBLP:conf/aaai/ParejaDCMSKKSL20} 
is implemented through a long short-term memory. However, like most GCN-based methods, the node aggregation of EvolveGCN is realized by averaging the information of neighboring nodes, which may not effectively take into account the importance of different neighboring nodes. To this end, we propose a new matrix computation method to update node representations based on the degree of neighboring nodes.

\section{Proposed Approach}
\begin{figure*}[t]  
	   		\includegraphics[width=18cm]{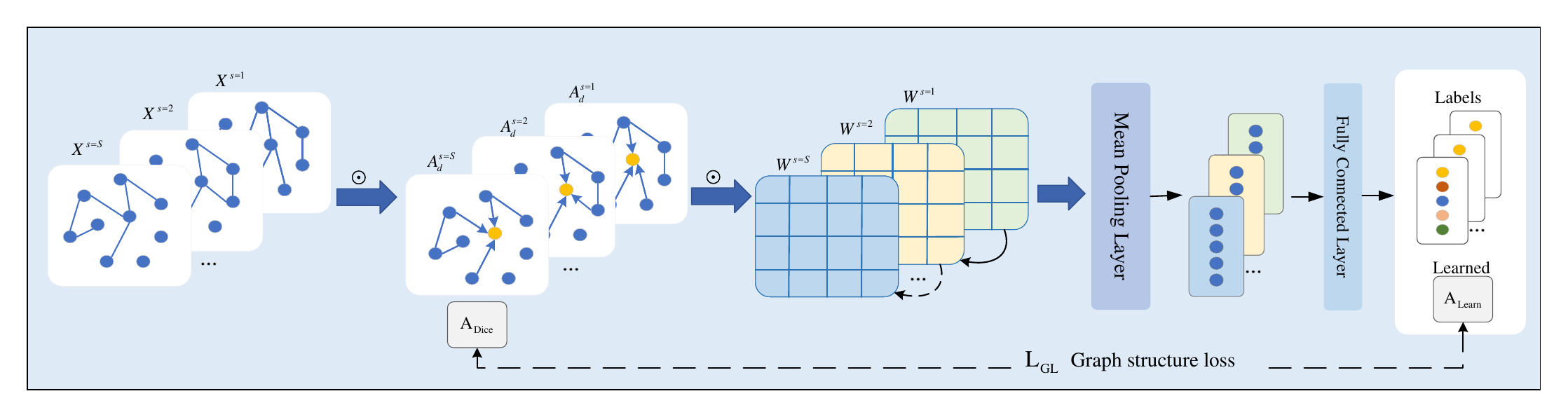}   
 \caption{Overview of our proposed model. The feature matrix $\mathbf{X}^{s}$, the graph structure (i.e., adjacency matrix $\mathbf{A}^{s}$), and the weight matrix $\mathbf{W}^{s}$ of the graph are dynamically changed with different audio segments $s$. Among them, the graph structure is mainly obtained by our proposed matrix calculation method ($\mathbf{A}_{Dice}^{s}$), and combined with the graph structure loss and classification loss to jointly optimize the graph structure. }
	\label{model_2}  
\end{figure*}

This section elaborates on our proposed architecture, which consists of three parts: the graph construction, the computation of relations between nodes, and the learnable graph convolutional layer. 
The overall framework of our model is shown in Fig.~\ref{model_2}.

\subsection{Graph construction}
Let $\mathcal{G} = \left \{ G^{1}, G^{2},\cdots, G^{S}   \right \} $ indicate a dynamic graph, different from the traditional methods that only construct a static graph for an entire sequence. $G ^ {s}$ is an observed graph specific to an audio segment $s$ $\left ( 1\le s\le S \right )$ and the structure of $G ^ {s}$ varies with different audio segments. $\mathcal{A} = \left \{ \mathbf{A}^{1}, \mathbf{A}^{2}, \cdots, \mathbf{A}^{S}   \right \}$ denote the structure of $\mathcal{G}$. 
More specifically, $G^{s} = \left \{ \mathcal{V}^{s},\mathcal{E}^{s},\mathbf{X}^{s}    \right \} $, where $\mathcal{V}^{s}=\left \{ v_{1}, v_{2}, \cdots, v_{m}\right \}$ indicate that each audio segment $s$ has $m$ nodes, $\mathcal{E}^{s}$ is an edge set at audio segment $s$, and $\mathbf{X}^{s}$ is a feature matrix for all nodes at audio segment $s$.
Given a dynamic graph $\mathcal{G}$, the goal of graph representation learning is to study a function $f: \mathbb{R}^{m\times m}\to \mathbb{R}^{m\times q} $ for each $G^{s}$ in $\mathcal{G}$. Specifically, based on the function $f$, the given $\mathcal{G}$ can be output as low-dimensional representations, e.g., $\mathcal{H}= \left \{ \mathbf{H}^{1},\mathbf{H}^{2},\cdots, \mathbf{H}^{S}   \right \}$, where $\mathbf{H}^{s}\in \mathbb{R}^{m\times q} $.

\begin{figure}[t]  
	\centering    
		\includegraphics[width=5.2cm]{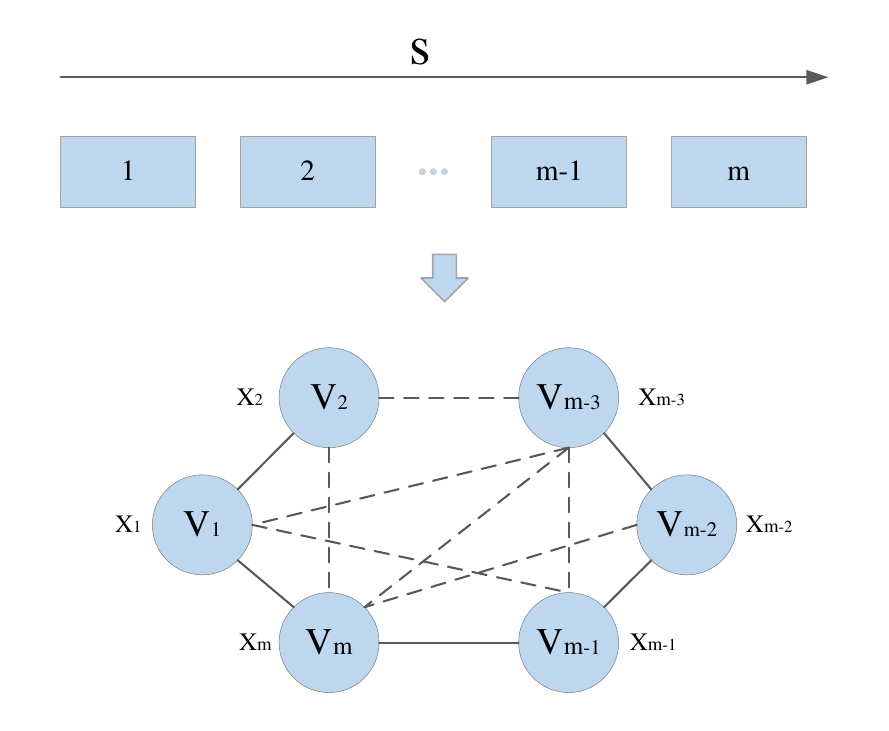}   
  \caption{Graph Construction. Given an audio segment $s$, frame it in 25ms steps to obtain $m$ frames. Then follow the frame-to-node transformation to get $m$ nodes. The list of edges is determined by connecting neighboring nodes and randomly connecting distant nodes. The relationship between nodes is jointly determined by the proposed matrix computation method and learning during the training. Each node has a node feature vector $x_{i}$ associated with it.}
	\label{cons}  
\end{figure}

Our graph construction follows the frame-to-node transformation, as shown in Fig.~$\ref{cons}$.
A frame represents a 25ms audio. To encode the temporal information, neighboring nodes (frames) need to be connected. 
Meanwhile, to aggregate the information of distant nodes, we also randomly connect the nodes. The $a_{ij} \in \mathbf{A}^{s}$  represents the weight corresponding to the edge $e_{ij} \in \mathcal{E}^{s} $ between $v_{i} $ and $v_{j}$  nodes. Note that the graph structure is not naturally defined here, i.e., the elements in $\mathbf{A}$ are unknown. The common methods to determine the $\mathbf{A}$ include cosine similarity function \cite{DBLP:conf/icassp/HuHWJM22}, manual definition \cite{DBLP:conf/icassp/ShirianG21}, and distance function \cite{DBLP:journals/inffus/ZhuMYZ22}. However, these may lead to a sub-optimal graph \cite{DBLP:journals/corr/abs-2303-02665}. Therefore, we propose a new matrix method to initially the $\mathbf{A}$ and optimize the graph structure during the training by combining graph structure loss and classification loss. This loss function will be presented in Subsection \ref{loss}. 

\subsection{A novel matrix computation}
Calculating the similarity between each node and its neighboring nodes is crucial in graph-based analysis. Dice similarity is a well-known node similarity measurement function \cite{DBLP:journals/pr/XieGWY18}. Since Dice similarity is calculated directly based on the network topology, it is relatively interpretable and saves computing resources. Given two nodes $v_{i}$ and $v_{j}$, the Dice similarity score is calculated as follows:
\begin{equation}
\setlength{\abovedisplayskip}{3pt}
\setlength{\belowdisplayskip}{3pt}
  \mathbf{S}_{Dice}\left ( v_{i} ,v_{j}  \right ) =  \frac{2Con\left ( v_{i} ,v_{j} \right ) }{N\left ( v_{i}  \right )+  N\left ( v_{j}  \right ) } ,
\label{Dice}
\end{equation}
where $Con\left ( v_{i} ,v_{j} \right )$ indicates the number of common neighbors. $N\left ( v_{i}  \right )$ and $N\left ( v_{j}  \right )$ denote the number of neighbors of $v_{i}$ and $v_{j}$ (not contain the node itself), respectively.
As can be seen from Eq.~(\ref{Dice}), when calculating the similarity between the nodes, the degree of neighboring nodes does not affect the final result. 

\begin{figure}[t]  
	\centering    
		\includegraphics[width=4cm]{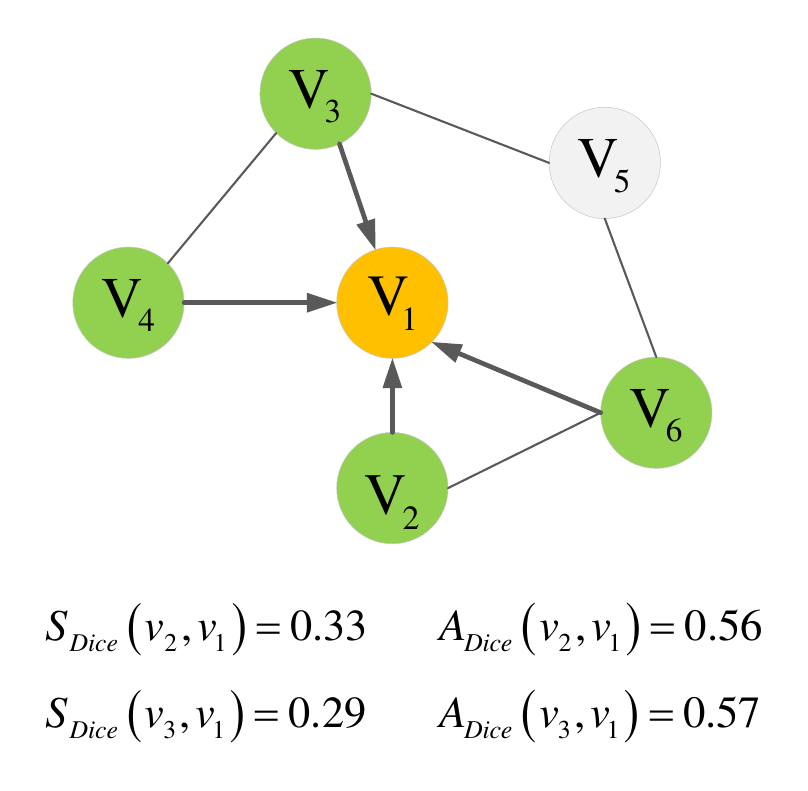}   
  \caption{Comparison of traditional Dice similarity $\left ( \mathbf{S}_{Dice} \right ) $ and proposed Dice similarity $\left ( \mathbf{A}_{Dice} \right ) $.  $v_{2}$  and $v_{3}$ have the same number of neighbors regarding the target node $v_{1}$; whilst the node with a larger degree ($v_{3}$) has a higher influence.} 
	\label{A_dice}  
\end{figure}

However, nodes with higher degrees are more likely to have more influence and higher weight values \cite{DBLP:conf/aaai/LiuN023}. For example, in the social networks of NBA players, players with more followers may bring more attention to their teams than players with similar physical conditions and skills. Take Fig.~$\ref{A_dice}$ as an example,
the common neighbors number of node $v_{2}$ with node $v_{1}$ is equal to the node $v_{3}$ with node $v_{1}$. The influence of node $v_{2}$ with smaller degree is $\mathbf{S}_{Dice}\left (v_{2},v_{1} \right) = 0.33$. However, the influence of node $v_{3}$ with larger degree is $\mathbf{S}_{Dice}\left (v_{3},v_{1} \right) = 0.29$, which is not expected in the real world. This is because graph-based methods usually update node representations by transferring and aggregating information from neighboring nodes. Thus, the higher-degree nodes may aggregate more information.

Therefore, we propose a new Dice matrix calculation method to solve the problem that the neighboring node degree has no positive effect on the target node. The proposed method is defined as follows:
\begin{equation}
Con\left (  v_{i} ,v_{j}  \right ) =  N\left ( v_{i}  \right ) \cap  N\left ( v_{j}  \right ),
\end{equation}
\begin{equation}
D\left (  v_{i} ,v_{j}  \right ) =  \frac{\left | N\left ( v_{i}  \right )\cup \left \{ v_{i}  \right \}    \right | }{\left | N\left ( v_{i}  \right )\cup \left \{ v_{i}  \right \}    \right | +\left | N\left ( v_{j}  \right )\cup \left \{ v_{j}  \right \}    \right | },
\end{equation}
\begin{equation}
\mathbf{A}_{Dice}^{s}\left ( v_{i} ,v_{j}  \right ) =  \frac{Con\left (  v_{i} ,v_{j}  \right )+  D\left (  v_{i} ,v_{j}  \right )}{\left | N\left ( v_{j}  \right )\cup \left \{ v_{j}  \right \}    \right | +1},
\label{new_dice}
\end{equation}
where $Con\left (  v_{i} ,v_{j}  \right )$ represents the number of neighbors shared by node $v_{i}$ and node $v_{j}$, and $D\left (  v_{i} ,v_{j}  \right )$ indicates the degree of neighbor node $v_{i}$ (the node itself is contained). According to Eq.~(\ref{new_dice}), as expected, the influence of node $v_{2}$ on $v_{1}$ and the influence of node $v_{3}$ on $v_{1}$ in Fig.~$\ref{A_dice}$ are $\mathbf{A}_{Dice}^{s}\left ( v_{2} ,v_{1}  \right ) =  0.56$ and $\mathbf{A}_{Dice}^{s}\left ( v_{3} ,v_{1}  \right ) =  0.57$. That is to say, the nodes with relatively large degrees have higher node weight values.

\subsection{Learnable graph convolution network}
\label{loss}
Traditional GCN layer usually takes the node feature matrix $\mathbf{X}\in \mathbb{R}^{m\times p}$ and the graph adjacency matrix $\mathbf{A}\in \mathbb{R}^{m\times m}$ as inputs to generate the node-level representation matrix $\mathbf{Z}\in \mathbb{R}^{m\times q} $. The GCN layer can be described as:
\begin{equation}
	\mathbf{H}^{\left ( l+1 \right ) }=\sigma \left ( \hat{\mathbf{A}}\mathbf{H}^{\left ( l \right ) }\mathbf{W}^{\left ( l \right ) } \right ) ,
	\label{normalized}
\end{equation}
where $\sigma$ is a $ReLU$ activation function that implements nonlinearity. $\hat{\mathbf{A}}=\mathbf{D}^{-\frac{1}{2} } \left ( \mathbf{A}+\mathbf{I} \right ) \mathbf{D}^{-\frac{1}{2} }$, $\mathbf{D}$ is a the degree matrix of $\mathbf{A}$, and $\mathbf{I}$ is a $m\times m$ identity matrix. $\mathbf{H}^{\left ( 0 \right ) } = \mathbf{X}$, $\mathbf{H}^{\left ( L \right ) } = \mathbf{Z}$, and $\mathbf{W}^{\left ( l \right ) }$ is the weight matrix for the $l^{th} $ layer. 

We present a new graph convolution structure for SER. It consists of the following two novel components:

$\bullet$ \textbf{Segment-specific graph convolution.}
For each audio segment $s$, we generate a node representation matrix $\mathbf{H}^{s} $. The graph convolution layer is represented as:
\begin{equation}
	\mathbf{H}^{s} = \sigma \left (\hat{ \mathbf{D}}_{s}^{- \frac{1}{2}  }{\mathbf{A}} ^{s} \hat{ \mathbf{D}}_{s}^{- \frac{1}{2}  }  \mathbf{X}^{s}\mathbf{W}^{s}  \right ) ,
 \label{G}
\end{equation}
where $\hat{\mathbf{D}}$ is a degree matrix of ${\mathbf{A}} ^{s}$ and ${\mathbf{A}} ^{s} = \tilde {\mathbf{A}}^{s} +\varphi \mathbf{A}_{Dice}^{s} +\mathbf{I}$. 
The $\tilde {\mathbf{A}}^{s}_{i,j}  = \left ( i-j \right ) ^{2} $, $\mathbf{A}_{Dice}^{s}$ is a novel matrix calculated by the proposed Eq.~(\ref{new_dice}), and $\mathbf{I}$ is an identity matrix.
$\mathbf{X}^{s}$ is a feature matrix and $\mathbf{W}^{s}$ is a weight matrix, where $\mathbf{W}^{0}$ is obtained by random and $\mathbf{W}^{s+1} =\sigma \left ( \mathbf{W}^{s} +{\mathbf{A}}^{s} *\mathbf{X}^{s}  \right )$.
Different from the traditional GCN that employs $\mathbf{A}+\mathbf{I}$ to guide the aggregate of node representations. We additionally add the $\mathbf{A}_{Dice}$ to further guide the aggregation and use a parameter $\varphi=0.6$ to control its contribution. Under the guidance of the new aggregation strategy, a higher-quality node representation can be obtained.


$\bullet$ \textbf{Learnable adjacency matrix ($\mathbf{A}_{Learn} $).}
We combine graph classification loss ($\mathcal L_{GC}$) and graph structure loss ($\mathcal L_{GL}$) to jointly optimize the graph structure during training. The $\mathcal L_{GC}$ is defined using the cross-entropy loss:
\begin{equation}
\setlength{\abovedisplayskip}{3pt}
\setlength{\belowdisplayskip}{3pt}
\mathcal L_{GC}= -\sum_{n=1}^{N}y_{n} \log_{}{\hat{y}_{n} } ,
\end{equation}
where $\hat{y}_{n}$ represents the prediction label for the $n$-$th$ sample. The $\mathcal L_{GL}$ is implemented as follows:
\begin{equation}
\mathcal L_{GL} = \underset{\text{graph structure loss}}{\underbrace{\lambda _{1}  e^{T}  \left ( \mathbf{A}_{Dice}  \odot  \mathbf{A}_{Learn}   \right ) e + \lambda _{2}  \left \|  \mathbf{A}_{Dice}  \right \|_{F}^{2}       } },
\end{equation}
where $\lambda _{1}$ and $\lambda _{2}$ control the proportion of these items respectively, $e$ is an all-ones vector, $\left \| \cdot
\right \|_{F}  $ indicates Frobenious norm, and $\mathbf{A}_{Dice}$ refers to Eq. (\ref{new_dice}).
The overall loss function is as follows:
\begin{equation}
\setlength{\abovedisplayskip}{3pt}
\setlength{\belowdisplayskip}{3pt}
    \min_{A_{l} ,p,\Theta } \mathcal L= \min_{A_{l} ,p,\Theta }\left [ \mathcal L_{GC}+  \mathcal L_{GL}   \right ] ,
\end{equation}
where $\Theta$ represents all learnable parameters on all graph convolution layers. Each term in the overall loss function $\mathcal L$ is differentiable, thus allowing end-to-end optimization.

\section{Experiments}
\begin{table*}[t]
\caption{Performance comparison between our model with latest (non-)graph models. Ablation experiments are constructing graph structures in different ways, such as the manually defined (\emph{\texttt{Binary}}), the distance-based (\emph{\texttt{Weighted}}), the learnable ($\mathbf{A}_{Learn}$), and the proposed node similarity matrix ($\mathbf{A}_{Dice}$). "-" represents no results provided in the references.}
\setlength{\abovecaptionskip}{0.2cm}
    \begin{subtable}[t]{0.49\linewidth}
    \vspace{-.2cm}
        \raggedright{
        \setlength{\abovecaptionskip}{0.2cm}
        \setlength\tabcolsep{3.4mm}
        \caption{Performance comparison on the IEMOCAP dataset.}
        \begin{tabular}{lcc}
		\hline
		\multicolumn{1}{l}{\multirow{2}{*}{Models}} & \multicolumn{2}{c}{IEMOCAP}                  \\ \cline{2-3} 
		\multicolumn{1}{c}{}                       & Acc(\%)         & F1(\%)   \\ \hline
         \multicolumn{3}{c}{\emph{Graph-based}}\\
        GCN (2017)\cite{DBLP:conf/iclr/KipfW17}  & 56.14         & -         \\ 
        PATCHY-Diff (2018)\cite{DBLP:conf/nips/YingY0RHL18}  & 63.23         & -           \\
         HSGCF (2023)\cite{DBLP:journals/kbs/WangDZLCHJ23}  & 65.13         & 65.18         \\
         DialogueGCN (2019)\cite{DBLP:conf/emnlp/GhosalMPCG19}  & 65.25         & 64.18          \\
         L-GrIN (2022)\cite{DBLP:journals/tmm/ShirianTG22}  & 65.50         & -             \\
        \hline
        \multicolumn{3}{c}{\emph{Non-graph based}}\\
        SpecMAE-12 (2023)\cite{DBLP:conf/icassp/SadokLS23}  & 46.70         & 45.90          \\
        CNN-LSTM (2019)\cite{DBLP:conf/interspeech/ParryPCLMBH19}  & 50.17         & -             \\
        DialogueRNN (2019)\cite{DBLP:conf/aaai/MajumderPHMGC19}  & 63.40         & 62.75          \\
       DualTransformer (2023)\cite{DBLP:journals/taslp/LiuKR23}  & 64.80         & 64.90          \\
        \hline
        \multicolumn{3}{c}{\emph{Adjacency matrix}} \\
        Ours (\texttt{Binary}) & 53.46 & 53.02\\
        Ours (\texttt{Weighted}) & 58.69 & 58.41\\
        Ours ($\mathbf{A}_{Learn}$) & 63.58 & 63.04\\
        Ours ($\mathbf{A}_{Dice}$) & 65.64 & 65.28\\
		\textbf{Ours} & \textbf{66.94} & \textbf{66.54}   \\ \hline
	\end{tabular}}
    \label{graph}
    \end{subtable}
    \hfill
    \begin{subtable}[t]{.48\linewidth}
        \vspace{-.2cm}
        \caption{Performance comparison on the RAVDESS dataset.}
        \setlength{\abovecaptionskip}{0cm}
        \setlength\tabcolsep{3.4mm}
        \begin{tabular}{lcc}
		\hline
		\multicolumn{1}{l}{\multirow{2}{*}{Models}} & \multicolumn{2}{c}{RAVDESS}                  \\ \cline{2-3} 
		\multicolumn{1}{c}{}                       & Acc(\%)         & F1(\%)   \\ \hline
         \multicolumn{3}{c}{\emph{Graph-based}}\\
         Synch-Graph (2020)\cite{DBLP:conf/aaai/Mansouri-Benssassi20}  & 42.60         & -             \\
          Esma et al.~(2019)\cite{DBLP:conf/ijcnn/Mansouri-Benssassi19}  & 45.10         & -             \\
           GCN (2023)  & 51.67         & 50.69             \\
          TSP-INCA (2021)\cite{DBLP:journals/kbs/TuncerDA21}  & 53.00         & 57.00             \\
          Riccardo et al (2022)\cite{DBLP:conf/icpr/FranceschiniFBC22}  & 58.50         & 57.00             \\
        \hline
        \multicolumn{3}{c}{\emph{Non-graph based}}\\
        SpecMAE-12 (2023)\cite{DBLP:conf/icassp/SadokLS23}  & 52.20         & 52.00          \\
        CNN-LSTM (2019)\cite{DBLP:conf/interspeech/ParryPCLMBH19}  & 53.08         & -          \\
         VGG-Transformer (2023)\cite{DBLP:journals/mta/GhalebNA23}  & 61.60         & -          \\
          GResNet (2019)\cite{DBLP:journals/mta/ZengMPY19}  & 64.48         & 63.11          \\
        \hline
        \multicolumn{3}{c}{\emph{Adjacency matrix}} \\
        Ours (\texttt{Binary}) & 51.67 & 50.69\\
        Ours (\texttt{Weighted}) & 51.46 & 50.59\\
        Ours ($\mathbf{A}_{Learn}$) & 64.10 & 63.72\\
        Ours ($\mathbf{A}_{Dice}$) & 65.69 & 65.34\\
		\textbf{Ours} & \textbf{67.50} & \textbf{67.05}   \\ \hline
	\end{tabular}
    \label{non-graph}
    \end{subtable}
    \label{result}
\end{table*}

\subsection{Database}
The \textbf{IEMOCAP} database contains 12 hours of data. To keep consistent with previous studies, only four emotions are used. We utilize the OpenSMILE tool to extract features from the audio. For each sample, we use a sliding window of length 25ms (with a step length of 10ms) to locally extract the low-level descriptors (LLDs), such as signal strength, power spectral density, and base frequency. Each speech sample contains 120 nodes, where each node represents an (overlapping) audio frame of length 25ms. 
The \textbf{RAVDESS} database includes 1,500 speech samples, performed by 24 professional actors (12 females and 12 males). Each actor simulated eight different emotional states. We use the Fourier transform to convert the speech signal into frequency domain representations. Then, we extract frequency domain features, such as the Mel frequency cepstrum coefficient (MFCC), where the sampling rate is 22,050 Hz and the number of MFCCs is 40. Each audio sample contains 40 nodes, where the node corresponds  to a 25ms audio frame. Appropriately increasing the number of MFCCs can help perceive changes in sound signals at different frequencies, and these detailed features may play a positive role in emotional sensitivity. The reason for extracting MFCC features from the RAVDESS dataset is that it provides representative spectral information.

\subsection{Implementation Details}
we trained the model to have up to 1000 epochs, with 150 iterations per epoch, and used an early stop strategy. The batch size of the model was set to 64.
Meantime, the RAdam optimizer with a learning rate of 0.001 was employed and we set the decay rate to 0.5 after every 150 epochs. 
We conducted experiments on GeForce GTX 3090 Ti, NVIDIA-SMI 460.39, and CUDA Version 11.2 GPUs, and obtained the final experimental data by using the 10-fold cross-validation averaging method.

\subsection{Results and Analysis}
\textbf{Comparison with SOTA methods.}
Table \ref{result} presents all results on two datasets.
On the IEMOCAP, compared with the graph-based SOTA approaches, the performance of our model seems only slightly better (1.44\% of accuracy absolutely) than L-GrIN \cite{DBLP:journals/tmm/ShirianTG22}, but the adjacency matrix $\mathbf{A}$ of L-GrIN is obtained by a distance function, which tends to locally optimal graphs during the training.  
When compared with other SOTA non-graph methods, we find that our method outperforms popular Transformer-based methods. This is mainly attributed to the fact that we connect neighboring nodes while randomly connecting distant nodes for information transfer. The relationship between the nodes is calculated by the proposed matrix calculation method to guide the aggregation of node information, thus capturing the long-distance information. 
Table \ref{result}(b) also shows that our model performs better than the graph-based and non-graph-based approaches on the RAVDESS database. For example, compared with the classic GCN model, our model has great performance advantages (15.83\% of accuracy absolutely). 
This indicates that the proposed matrix computation method can alleviate the sub-optimal graph in traditional GCN caused by averaging the information of neighboring nodes.

\textbf{Ablation experiment.}
We perform ablation studies on two datasets, including the following variations of our model: \texttt{Binary}: manually defining the structure of the graph, the matrix $\textbf{A}$ only contains $0$ and $1$; \texttt{Weighted}: the relationship between nodes is determined based on the distance; 
$\mathbf{A}_{Learn}$: the structure of the graph is learned during the training; $\mathbf{A}_{Dice}$: the loss function only contains the proposed matrix computation method.
The results of the study are displayed in the ``\emph{Adjacency matrix}'' of Table \ref{result}. We have the following observations.

Firstly, both \texttt{Binary} and \texttt{Weighted} determine the relationship between nodes through customized form, which makes them easy to form the local optimal graphs during the training process, so the performance is the worst.
Secondly, the proposed matrix calculation method $\mathbf{A}_{Dice}$ has enhanced performance compared to the learnable $\mathbf{A}_{Learn}$, which shows that our method can effectively distinguish the significance of different neighboring nodes.
Finally, we optimize the structure of the graph by jointing the graph structure loss and classification loss to achieve the best performance. This indicates that the proposed matrix computation method can update the node representation and that the structure of the graph can be continuously optimized during the training process.

\section{Conclusion}
We propose an adaptive graph representation learning method based on dynamically evolved graphs rather than the traditional methods that only construct a static graph for
an entire sequence. Our graphs are consecutively constructed on a series of subsequences segmented by a sliding window. To compute the edge weights, we propose a new matrix calculation method that updates the node representations based on the degree of neighboring nodes.
Moreover, we combine the graph structure loss and classification loss to jointly optimize the graph structure during the training.
In the future, we will simultaneously consider combining structural similarity with feature similarity to jointly measure the similarity of nodes, and work on multimodal data with graph structures.

\newpage
\bibliographystyle{IEEEbib}
\bibliography{myreference}
\end{document}